\documentclass[11pt,a4paper]{article}

\renewcommand{\author}{Emil Khalisi}
\newcommand{\titel}{The Solar Eclipses of the Pharaoh Akhenaten}
\newcommand{\version}{Version 1.24}
\renewcommand{\date}{20th July 2020}

\usepackage{hyperref}
\hypersetup{pdfauthor={Emil Khalisi}}
\hypersetup{pdftitle={\titel}}
\hypersetup{pdfsubject={Online publication, \version of \date }}
\hypersetup{pdfkeywords={Astronomical Dating, Chronology, Egypt, Akhenaten,
      Solar Eclipse, Amarna, New Kingdom, Dynasty XVIII}}

\usepackage[T1]{fontenc}        
\usepackage{mathptmx}           
\usepackage{pifont}             
\usepackage[scaled=0.92]{helvet} 
\usepackage[british]{babel}     
\usepackage{amsmath}            
\usepackage{microtype}          

\usepackage{multicol}           
\usepackage{multirow}           
\usepackage{eurosym}            
\usepackage{units}              
\usepackage{setspace}           

\usepackage{calc}
\usepackage[a4paper]{geometry}  
\usepackage{scrextend}
\changefontsizes{10pt}
\usepackage{titlesec}
\titleformat*{\section}{\large\bfseries}
\titleformat*{\subsection}{\normalsize\bfseries}

\usepackage{graphicx}           
\usepackage{float}              
\usepackage{caption}            
\usepackage{fancyhdr}
\renewcommand{\headrulewidth}{0.4pt}
\usepackage{colortbl}             
\definecolor{grey20}{RGB}{208,208,208}

\begin{document}


\fancyhead{}
\fancyhead[LO]{%
   \footnotesize \textsc{In original form published in:} \\
         Habilitation at the University of Heidelberg
}
\fancyhead[RO]{
   \footnotesize {\tt arXiv: 2004.12952 [physics.hist-ph]}\\
   \footnotesize {v2: \date }%
}
\fancyfoot[C]{\thepage}

\renewcommand{\abstractname}{}

\twocolumn[
\begin{@twocolumnfalse}

\section*{\centerline{\LARGE \titel }}

\vspace{\baselineskip}
\begin{center}
{\author \\}
\textit{69126 Heidelberg, Germany}\\
\textit{e-mail:} \texttt{ekhalisi@khalisi.com}\\
\end{center}

\vspace{-\baselineskip}
\begin{abstract}
\changefontsizes{10pt}

\noindent
\textbf{Abstract.}
We suggest an earlier date for the accession of the pharaoh
Akhenaten of the New Kingdom in Egypt.
His first year of reign would be placed in 1382 BCE.
This conjecture is based on the possible witness of three annular
eclipses of the sun during his lifetime:
in 1399, 1389, and 1378 BCE.
They would explain the motive for his worship of the sun that left
its mark on later religious communities.
Evidence from Akhenaten's era is scarce, though some lateral
dependencies can be disentangled on implementing the historical
course of the subsequent events.

\noindent
\textbf{Keywords:}
Solar eclipse,
Astronomical dating,
Akhenaten,
New Kingdom,
Egypt.


\end{abstract}

\vspace{\baselineskip}
\centerline{\rule{0.8\textwidth}{0.4pt}}
\vspace{2\baselineskip}

\end{@twocolumnfalse}
]



\section{Introduction}

The flourishing time of the 18th to 20th dynasty of the Egyptian
pharaohs, the so-called ``New Kingdom'', is not well established.
Traditionally it is placed roughly between 1550 and 1070 BCE.
In the public awareness this era of ancient Egypt is known best,
since most people associate with it the ``classic pharaonic
etiquette''.
Memphis near today's Cairo was the administrative center in the
very old times, while Thebes about 650 km farther to the south
remained an important residence of the monarchs.
Thebes was restored as capital in a re-unified Egypt with the
commencement of the 18th dynasty.

One of the most prominent pharaohs of the 18th dynasty was
Akhenaten (Amenhotep IV).
The interval of his reign has been subject to much debate.
Depending on the method of reckoning, the accession of the
illustrious pharaoh is assumed somewhere in the middle of the 14th
century BCE.
Our attempt will be to investigate the year of his accession by
some kind of guesswork.
First, we briefly recap the political circumstances and outline
important stages in Akhenaten's life.
In particular, we ask for his reasons to carry out the radical
changes in the religious belief and to abandon the polytheistic
customs in favour of a monotheistic worship of the sun.
We rest our answer on his experience of solar eclipses and present
some striking events.
Our hypothesis seems speculative, but we claim that the timeline
agrees with other incidents in the subsequent history better than
with the current doctrine.
Connections to the Hittites provide the most important linkage
supporting an early time slot for Akhenaten.

A crucial element when attributing historical eclipses to a certain
geographical region is the clock-time correction, $\Delta T$, that
shifts the eclipse path in accordance with the deceleration of the
earth's rotation.
That error accumulates to more than 9.0$\pm$0.5 hours for the
period of time considered here.
Though there are many reasons to refrain from this method for dates
before 700 BCE, we argue that the average $\Delta T$ is sufficient
to satisfy the timeline.
The exact position of the central tracks is not required to suit
our revised course of the historical cornerstones.


\section{Worship of the Sun}

\enlargethispage{-1ex}

The adoration of the most important luminary in the sky played a
central role for the old Egyptians, in religion as well as in
culture.
That civilisation was the first one to arrange its calendar after
the sun and to disengage from auxiliary solutions with lunar leaps
as undertaken in Babylon or China.
All the more it seems surprising that there are no obvious mentions
about obscurations of their major object of cult.
A sudden darkening of the sun must have caused a horrific shake of
the belief.
In the same manner eclipses of the moon must have been noticed,
but we do not find anything about worries, dramatic uproars, or
attempts at interpretation.
A complete ignorance of these natural phenomena appears
incomprehensible.

One possible explanation for the absence of eclipse accounts may
lie in the governmental structure.
The Egypts lived in an absolute authoritarian society.
Only few people were capable of writing, and they were strictly
forbidden to announce malfunctions within the system that would
equal a dare over the godlike pharaoh.
The ban on speaking affected any kind of turmoils, social tensions,
shortages, and, even more, religious affairs.
If something should ever leak, then probably in a highly cryptic
way --- neither mention of names, nor dates, nor obvious details.
The episode of the 10th pharaoh of the 18th dynasty, Akhenaten,
is an example for a far-reaching disharmony in the state.

His father, Amenhotep III, favoured the god Aton (Sun) over all
other deities and supported the rites.
Probably he aimed at a smart change in politics to counter the
local tradition of Amun (creationism) which was sustained by the
powerful priesthood.
His son Akhenaten became pharaoh at an estimated age of 18 to 22
years \cite{habicht-etal_2016}.
He executed drastic changes, in his fifth year, by elevating the
sun over all existing gods
(other historians speak of the third year of reign).
The sun would take over the sole universal mastery in the Egyptian
pantheon.
Then, he appointed himself its descendant and changed his name:
``Akhen-Aten'' means ``the one serving the sun''.
As a pharaoh he deprived the priests off their power and ordered
to stop the former Amun rites.
He tempestuously promoted sculptures, portraitures, and amulets
with sun-related features.
Moreover, he dedicated a new capital to it:
Achet-Aton, today known as Amarna.
After a rapid construction he moved there in his 8th year.

The objectives of these acts are quite sketchy, and the background
is controversial among Egyptologists.
It remains an open question why Akhenaten took the glorification
of the sun to extremes.
It seems another case for the stock of legends.
In any case, the reorganisation of the religious belief must have
run into opposition by the priests.
Historians call this the ``Amarna Period''.
It is likely that a few Jews, who were working as slaves in the
Egyptian exile at that time, were inspired by Akhenaten's new
monotheistic mindset and incorporated these ideas into their own
philosophy \cite{assmann_1997}.
The Old Testament emerging very much later was re-written in order
to retroactively adjust a primordial pagan socket religion with
the new thoughts evoked by Akhenaten.
Maybe these workers were even constructing Amarna, and,
hence, became acquainted with the subject.

The pharaoh died in his 17th year of reign, and the circumstances
of his death are obscure.
His successor, Smenkhkare, pops up as an enigmatic figure that
seems to conceal someone else.
It is not clear whether a man or a woman hides behind that name,
from who descending, or epitomising even two persons.
Akhenaten's only son, Tutankhamun, was an infant at an age of
four when inaugurated as pharaoh \cite{fru_wente-siclen}.
An age of 8 or 9 is also conceivable.
However, Tutankhamun's kinship is not conclusively clarified, as
genetical analysis proves intricate patterns between the pharaohs.
The complexity of the family relationships is reviewed, e.g., by
Michael Habicht \cite{habicht-etal_2016}.

It is said that Tutankhamun withdraw many changes introduced by
his father, most likely upon the pressure of the priesthood.
For them, it would be easy to bring an underaged to restore the
religious practice back to the old Amun cult.
The birthname of the child pharaoh was originally
``Tutankh-Aten'', like all his sisters bearing the syllable for
the sun, but he altered it for the sake of the old gods and former
rites.
In Tutankhamun's second year the capital was moved back to Memphis,
and Thebes lost significance even more.
He died after nine or ten years in power
at a maximum age of 20.
After him the worship of the sun was to be liquidated and all
traces erased.
Tutankhamun never stood out politically in any way.
For the history of the country he was completely irrelevant.
His present-day fame is just based on the sensational discovery
of his undamaged tomb in 1922.

%
\fancyhead{}
\fancyhead[CE, CO]{\footnotesize \itshape E.\ Khalisi (2020): \titel}
\renewcommand{\headrulewidth}{0pt}

\begin{table}[t]
\caption{Suggested reign of Akhenaten (selection).}
\label{tab:akhenaten}
\centering
\begin{tabular}{l@{ -- }rlc}
\hline
\rowcolor{grey20}
\multicolumn{2}{c}{\cellcolor{grey20}{Time [BCE]}} & Method & Ref. \\
\hline
1382 & 1367 & Moon visibility + tables & \cite{fru_huber_2011} \\
1380 & 1365 & {\it Cambridge Ancient History} & \cite{fru_kelley-milone} \\
1377 & 1356 & Synchronisms with Hatti & \cite{rice_2003}   \\
1367 & 1350 & Links to Assyria & \cite{fru_kelley-milone} \\
1360 & 1343 & Genealogical relationships & \cite{dobson-hilton} \\
\multicolumn{2}{l}{1359$\pm$11 (3$\sigma$)} 
            & Radiocarbon dating & \cite{fru_ramsey_2010} \\
1356 & 1340 & Geo-political links & \cite{fru_gertoux_2013} \\
1353 & 1336 & Egyptological research & \cite{hornung_2006} \\
1351 & 1334 & Crosslinks to Babylon & \cite{eder_2004} \\
1350 & 1334 & Dynasty lists & \cite{fru_wente-siclen} \\
1349 & 1333 & Anthropological examination & \cite{habicht-etal_2016} \\
1340 & 1324 & Archaeology & \cite{w-de} \\
\multicolumn{2}{c}{$<$ -1300} & Dendrochronology & \cite{fru_kelley-milone} \\
\end{tabular}
\end{table}

Akhenaten's reign is situated by historians at the middle of the
14th century BCE.
Depending on the method of analysis, various time slots for his
rulership are proposed (Table \ref{tab:akhenaten}).
For the entire 18th dynasty there are rumours of co-regencies that
would complicate the reckoning.
In the case of Akhenaten the claims fluctuate between 8 and 12
years of co-regency, however, they are also vehemently rejected
by other historians.

Manifold attempts to set absolute time limits for the 18th dynasty
turn out unsteady.
Information is patched together from various disciplines, as one
tries to generate a holistic picture.
Historians tend to use a ``short chronology'' for their needs, but
the ``long chronology'' offers better results from the astronomical
point of view \cite{fru_huber_2011, finsternisbuch}.
The lifetime of Akhenaten casts shadows over the dates for the
Hittite king Mur\v{s}ili II, the ``Dakhamunzu'' episode, and the
Battle of Kadesh several decades later.


\section{Previous Suggestions for Eclipses}

In the variety of works on dating Akhenaten, we restrict ourselves
to eclipses.
Studies of similar kind were presented by the engineer and
historian William McMurray \cite{fru_mcmurray},
the archaeoastronomer G\"oran Henriksson \cite{fru_henriksson_2007},
and the astrophysicist Giulio Magli \cite{magli_2013}.
Another analysis was provided by Peter Huber who thoroughly
synchronised various moon dates with pillars from history
\cite{fru_huber_2011}.

%
Reviewing McMurray briefly, he expressed the idea that an image on
a grave wall, discovered in Amarna, depicts an allusion to a solar
eclipse.
The tomb is assigned to the high priest Meryra I and is categorised
as a major private one from the 18th dynasty.
Meryra's task was to care for the new Aton service launched by the
pharaoh.
The tomb is estimated at the 9th or 10th year of Akhenaten's reign.

\begin{figure}[t]
\includegraphics[width=\linewidth]{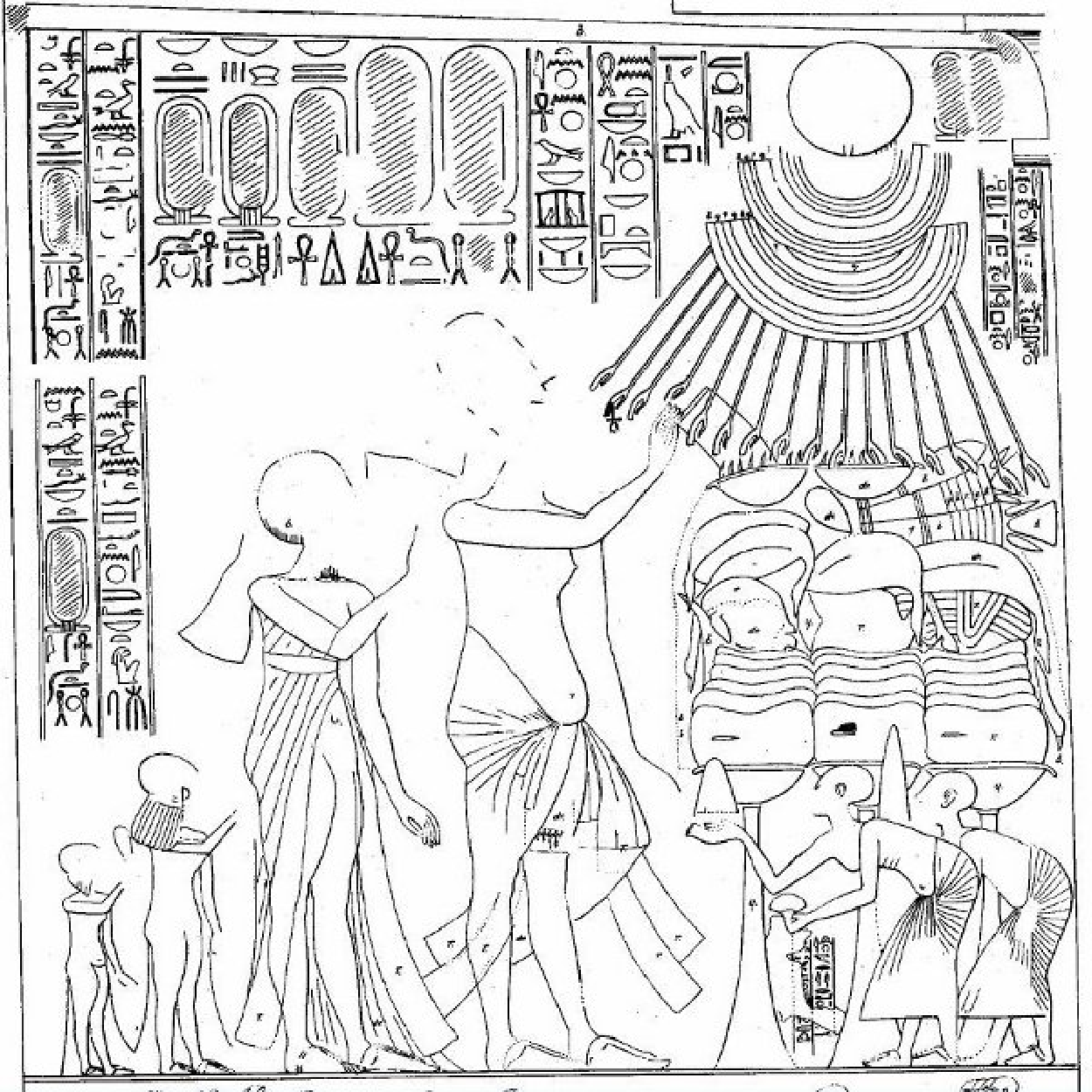}
\caption{Replica of a worship scene by Akhenaten and his wife in
    the tomb of Meryra I \cite{fru_mcmurray}.}
\label{fig:meryretomb}
\end{figure}

\begin{figure}[t]
\includegraphics[width=\linewidth]{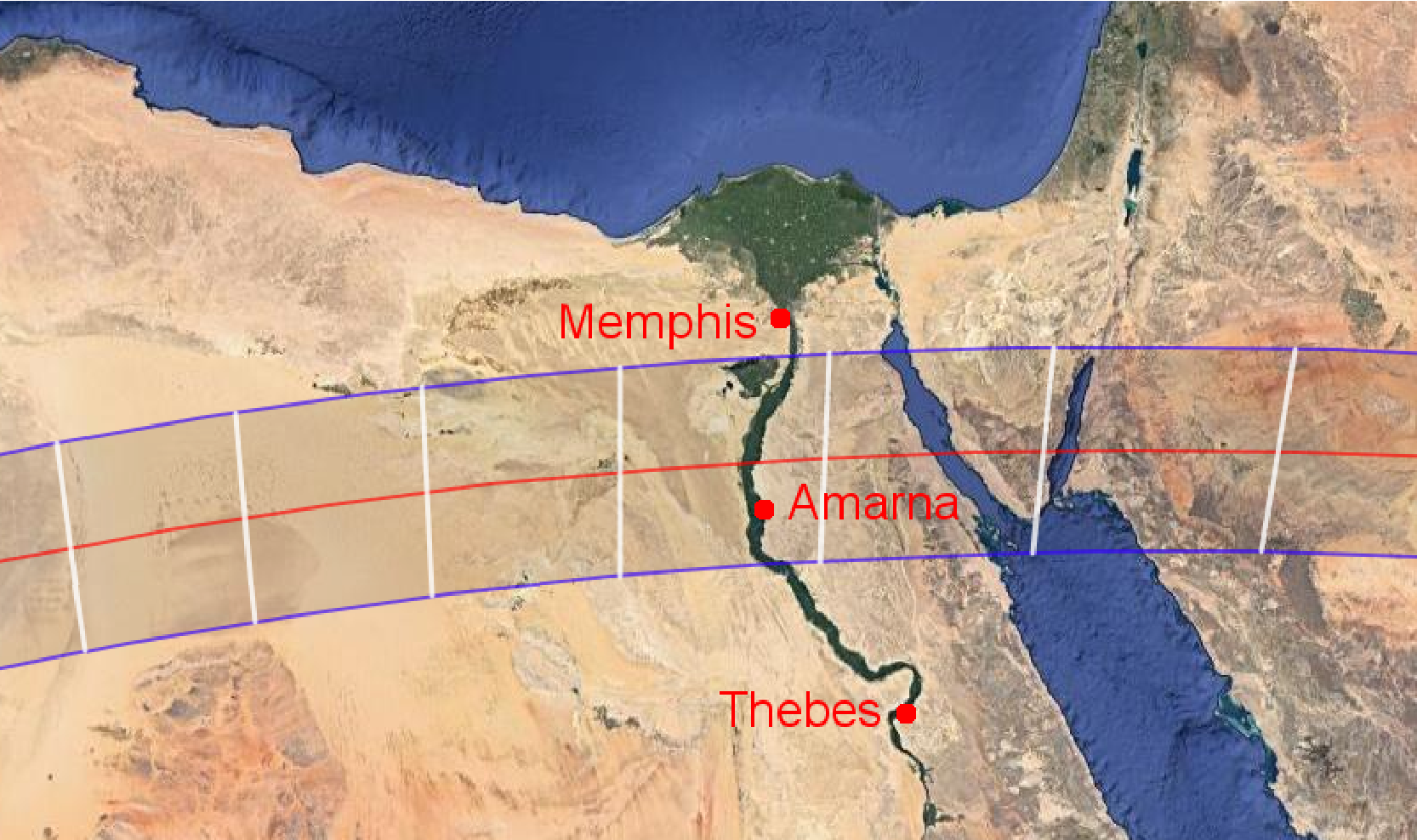}
\caption{Total solar eclipse of 1338 BCE \cite{espenak}.}
\label{fig:echnaton-bc1338}
\end{figure}

Among the countless sketches showing the pharaonic couple beneath
the beamy sun, McMurray selected one with arcs
(Figure \ref{fig:meryretomb}).
These arcs are still not understood.
The interpretations cover a wide range of suggestions:
collars, clouds, upside-down rainbows, or atmospheric halos.
Also, flying shadow bands were proposed appearing shortly before
the totally eclipsed sun.
Resting his arguments on the latter option, McMurray pleads in
favour of the total solar eclipse of 14 May 1338 BCE
(Figure \ref{fig:echnaton-bc1338}).
The track passed a few kilometers south of Cairo (Memphis) at
noonday at 2 p.m.,
and it is quite insensitive to the clock-time correction
$\Delta T$ because of its course in west-to-east-direction.
The magnitude was 0.991 in Memphis and 0.943 in Thebes, respectively.
That event would have inspired Akhenaten to build the new residence
halfway between these cities where totality was actually achieved.
Thus, it would be no factor of chance that the complex was raised
in a deserted area.
Akhenaten ascended the throne in 1340 BCE, i.e.\ two years before
the eclipse, because of special festivals celebrated at full moon,
and after the eclipse those heretic changes were initiated.
McMurray asserts a consistency with appropriate inscriptions on
temples.

%
The same eclipse was favoured by Giulio Magli upon consideration
of architectural alignments of temples in Amarna \cite{magli_2013}.
The whole urban plan was fixed depending on the location of the
wadi and tomb in such way that ideal straight lines connect the
tomb with all main buildings.
Even if a few things may stretch some things too far, as the
author admits, there are boundary stelae arranging symmetrical
spaces in which the tomb itself is the focus of attention.
Amarna would have been constructed before that total eclipse of
1338 BCE, and the unexpected incident would have taken place at
the end of Akhenaten's reign.
It was interpreted as a negative sign:
a spell was cast by the old gods on the sun,
and the priests would have ``welcomed'' that as the call to
return to the former status quo.
Akhenaten was replaced, after his mysterious death, by Smenkhkare
first and then by the child-king Tutankhamun.

So, in contrast to McMurray who proposed an active impetus
received by that eclipse (construction of Amarna), Magli prefers
a passive interpretation:
the site was evacuated because of it.
Additional evidence for this hypothesis would come from the
celestial circumstances at the instant of totality:
it occurred in that part of the sky which lies between the star
constellations of Gemini and Taurus.
The sun is closest to the celestial region pertaining the
Osiris-Orion, as defined by the old Amun religion.
According to Magli, the eclipse ``communicated'' the divine sign
for the reversal of policy.

%
A different eclipse was brought forward by G\"oran Henriksson
\cite{fru_henriksson_2007}.
He selected six partial and annular eclipses from a large sample
and constrained them to the horizon, because small obscurations
will be noticed there much better than in the glare of the day.
A partial eclipse can easily escape attention if the magnitude is
smaller than $\approx$0.7 \cite{finsternisbuch}.
Henriksson showed that the annular eclipse of 5 July 1378 BCE was
very conspicuous, especially, in Luxor at the king's residence
near Thebes (Figure \ref{fig:echna-tripel}).
The sun would have risen with a circular black ``hole'' inside.
It happened in the sixth year of Akhenaten's reign and would be
the cause for his name change.

Three years later, on 3 May 1375 BCE, another eclipse
(partial, mag= 0.795 in Luxor) would have triggered the pharaoh's
definite decision to construct Amarna.
Here, Henriksson deviates from the majority that the idea took
shape before the fifth year.
And, furthermore, a third eclipse might have induced his son,
Tutankhamun, for the turnaround:
27 October 1356 BCE (partial, mag = 0.852 in Amarna).
The priests would have urged to abandon Amarna, and the child
pharaoh went with it.


\section{An Eclipse Triple?}
\label{ch:triple}

Following the chronology by Peter Huber \cite{fru_huber_2011},
only three years are suitable for the time of Akhenaten's
accession: 1353, 1378, and 1382 BCE.
The years are based on the visibility of lunar crescents that led
to a most probable calendar round.

\begin{figure*}[t]
\includegraphics[width=\linewidth]{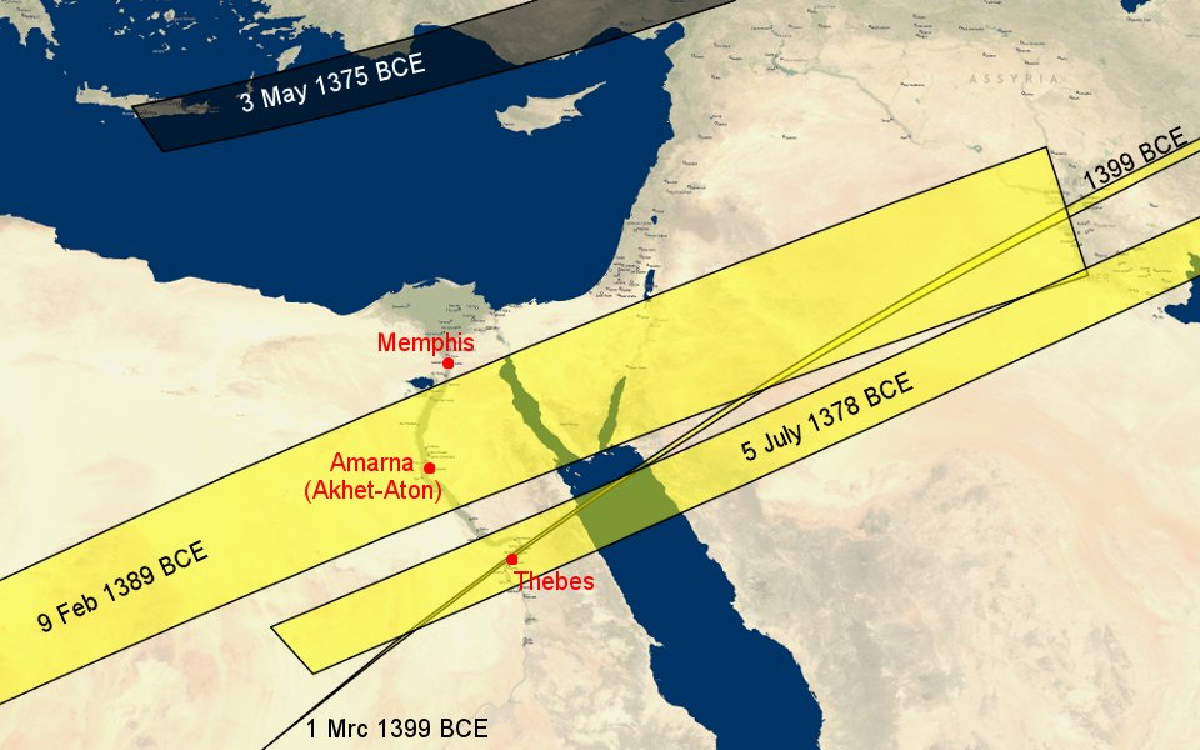}
\caption{Akhenaten could have witnessed three annular eclipses in
    his lifetime.
    The total eclipse of 1375 BCE (grey) was partial in Egypt and
    occurred at sunrise.}
\label{fig:echna-tripel}
\end{figure*}

Although a solar calendar was used for indicating dates, the
everyday's life of the common people was guided by the phases
of the moon.
For keeping pace of the solar days with the lunar month, a
``psdntyw day'' was inserted, that operated like a double date,
if the visibility of the lunar phase fell behind.
This ``psdntyw day'' was introduced by Heinrich Brugsch in 1864,
and discussed by follow-up researchers, see
\cite{fru_huber_2011, parker_1957} for details.
Huber analysed such days on the basis of the visibility of the
lunar crescent at its conjunction with the sun and concluded that
only the aforementioned years are viable to Akhenaten's year 1.
Therein, the interval of 4 years between 1382 and 1378 BCE is
intriguing.

We checked the eclipses of the period from 1420 to 1320 BCE in
Espenak's \textit{Five Millennium Canon of Solar Eclipses}
\cite{espenak}.
Relying on the ancillary information from history that the
construction of the new capital started in Akhenaten's fifth year,
we presume a solar eclipse prior to that.
Other eclipses in the childhood of the pharaoh could also have
influenced his attitude towards worship, though such things stay
rather speculative.
Altogether, Akhenaten could have seen some impressive events
before he became regent.
His father, Amenhotep III, ruled for 38 years, and it is believed
that he died at an age of almost 50, when Akhenaten inherited the
throne \cite{hornung_2006}.

Indeed, we meet the solar eclipse of 1 March 1399 BCE first.
Being of hybrid type, it had a maximum path width of 10 km
(Figure \ref{fig:echna-tripel}).
For the average $\Delta T$, the track swept exactly over Thebes
which was itself close to the turning point from totality to
annularity (mag $\approx$ 0.996 at 4:30 p.m.).
We will never know what sight the blackened sun displayed to the
viewer, but one can imagine that Baily's Beads flickered around,
and the terrifying circle was surrounded by a reddish chromosphere
making the scene appear diabolical.
Whatever it looked like to the observer, if he stood on the central
path, the phenomenon would have created a stunning show of light.

If assuming 1382 BCE for the accession of Akhenaten, he must have
been a child $\approx3\pm$2 years of age at that spectacle in 1399.
His father would be pharaoh and surely not less astonished at the
strange sight as it was.
The father could have experienced another eclipse 11 years earlier,
on 25 September 1410 BCE:
also of annular type, right after sunrise (mag = 0.923 in Memphis
or 0.864 in Thebes).
It must have caught attention unless weather conditions impeded
the view.

Akhenaten, as a juvenile, might have experienced his second
eclipse ten years later, on the evening of 9 February 1389 BCE.
Again of annular type, the sun exhibited a luminous ring around
a black disk.
In such sinister state the sun set in Memphis, or partially in
Thebes (mag = 0.891).
Thus, two striking eclipses occurred before his regency.

Eleven years onward Akhenaten would have observed the third one,
now in his role as monarch:
5 July 1378 BCE, as suggested by Henriksson.
In Thebes it was seen (mag = 0.982) in the morning hours, and
in Memphis it was impressive, too (mag = 0.872).
This happened in his fourth year and might have inspired him to
construct Amarna.
Furthermore, another (fourth) great eclipse would have fallen into
his reign.
On 3 May 1375 BCE, at sunrise, the solar disk climbed over the
horizon being partially obscured (mag = 0.854 in Amarna, and 0.796
in Thebes).

Akhenaten would have witnessed three central eclipses in his
lifetime, if he had stayed, by chance, in Memphis and Thebes at the
respective days.
If this was not the case, and he remained throughout in Thebes
before moving to Amarna, two of the eclipses are impressive enough
to leave behind a strong impression.
Only very few people get the chance of sighting more than one
central eclipse, the majority even none.
Such experiences do affect the attitude, and being a child all the
more.
Our claim now is that the repeated observation of these oddities
laid the foundation for Akhenaten's worship of the sun.
A boundary stone was found in Amarna bearing an inscript stating
that \dots
``he would never cross the southern nor northern border of his
city but wants to be buried towards the sunrise after a million
jubilees of regency'' \cite{schloegl_2008, w-de}.

Now, let us check whether or not the years of the three eclipses
would fit to the course of history.


\section{Discussion}

Solar eclipses strongly depend on the location of the observer.
When they are historical, they will additionally be subject to the
clock-time error, $\Delta T$, that denotes the difference between
the strictly uniform time scale (Ephemeris Time) and the civil time
(Universal Time).
The difference is based on the long-term slowdown of the Earth's
rotation due to tidal friction, but other geophysical causes apply
as well.
The value of $\Delta T$ can be approximated by a parabola
$\sim 32.5 \cdot t^2$ with $t$ in centuries before 1800.
This issue has been investigated in many publications, see the
extensive research by Richard Stephenson \cite{histeclipses} or
the most recent overview by the author \cite{finsternisbuch}.

Avoiding repetitions, we just make mention that eclipse paths are
very sensitive to $\Delta T$ when going far back in time.
In the case of the hybrid eclipse of 1399 BCE, it amounts to
32,946$\pm$1,593 s \cite{espenak}.
However, there exist other models producing a slightly different
value, but all of them obey the parabolic formula.
Though the error in $\Delta T$ is decisive, we emphasise that it
is the \emph{average} of this extrapolation being fully adequate
to reproduce the coherent scene as discussed here.
Furthermore, the exact position of the central zone relative to
the observer is of minor relevance.
A small shift of the geographical course of the eclipse to the
west or east would not change much of the individual sight at the
obscured sun,
but the choice of the correct eclipse for dating purposes will.

So, what reference points do we have, in general, in order to
select the proper event for Akhenaten?
Astronomy supplies the most accurate pegs, but for Egypt of the
pharaohs there are almost no celestial hints handed down.
There is nothing to be used as a pillar than the calendrical
system, whether it is linked to agriculture, lunar visibility, or
heliacal rising of Sirius.
Therefore, historians usually make a detour via the Hittites or
Babylonians, borrow a secured date from there, and start computing
``distances'' in time deploying crosslinks.
One of such kind is the so-called ``solar omen'' of Mur\v{s}ili II,
another one could principally be the Battle of Kadesh.
A third way is trickier:
reconstructing the timeline from the capture of Babylon after the
Hammurabi-Dynasty down to the political Dakhamunzu affair that
followed Akhenaten's death.
The latter way, however, is dependent on the choice of the
chronology (high, middle, low) for the ancient Near East.

\noindent
\begin{enumerate}
\item
For Mur\v{s}ili's solar omen, which is interpreted as a solar
eclipse in Anatolia, we refer to Chapter 9.7 of
\cite{finsternisbuch}.
Seven possible eclipses between 1340 and 1301 BCE are on the
shortlist for a closer inspection.
Many authors declare themselves in favour of the event of 1312 BCE.
After working backwards through the Dakhamunzu affair, they arrive
at accession years for Akhenaten around 1350 BCE, and they feel
done.
A motive for his worship of the sun is missing.
If these authors had adopted the year 1340 BCE for the solar
eclipse, as suggested by astronomers earlier, they would have
arrived 30 years further back in time, and the motive would light
up immediately --- see above.

\begin{table}[t]
\caption{Proposed years [BCE] for the Battle of Kadesh.
    The year (*) 1274 BCE is favoured by most historians.}
\label{tab:kadesh}
\centering
\begin{tabular}{rlc}
\hline
\rowcolor{grey20}
 Year & Method of evidence     & mentioned in \\
\hline
 1335 & Lunar dates & \cite{fru_huber_2011} \\
 1310/11 & Lunar dates & \cite{fru_huber_2011} \\
 1299 & Lunar dates & \cite{parker_1957} \\
 1296 & Lunar dates & \cite{parker_1957} \\
 1285 & Lunar dates & \cite{casperson_1988} \\
 1280 & ($\pm$12) Radiocarbon analysis & \cite{fru_ramsey_2010} \\
 1278 & Synchronisms to Assyria & \cite{fru_gertoux_2013} \\
 1274 & Lunar dates & (*) \\
(1271) & Lunar dates (unlikely) & \cite{parker_1957} \\
 1260 & Lunar dates & \cite{fru_huber_2011} \\
$<$1250 & Archaeological considerations & \cite{fru_huber_2011} \\
 1249 & Lunar dates & \cite{fru_huber_2011} \\
 1235 & Lunar dates & \cite{fru_huber_2011} \\
\end{tabular}
\end{table}

\item
The second support is obtained from the Battle of Kadesh fought
between the fifth successor of Tutankhamun, Ramesses II, and the
Hittites.
This military campaign is described extensively on many temple
walls, and it took place in the 5th year of Ramesses II.
The year 1274 BCE is entrenched in the minds of historians,
but how is it worked out? ---
Those, who prefer a sound practice, use astrochronology, again,
but this time the trail comprises some kind of ``island-hopping''.
First, one has to find the accession year of Ramesses II.
In his 52nd year, a lunar date ``II prt 27'' is recorded as being
``psdntyw day'', i.e.\ the first day of the lunar month when the
crescent was not visible \cite{parker_1957, casperson_1988}.
When performing detailed astronomical calculations, several years
turn out possible (Table \ref{tab:kadesh}).
From there, the other points in the life of Ramesses II would
emerge, and so does the Battle of Kadesh.
The mounting of further jigsaw pieces can continue.
The comparison of literature discloses much shakier results than
hoped for a reliable synchronisation.
The widespread year 1274 BCE proves far from certain.
In view of the computation by Peter Huber (visit his discussion to
Table 4 in \cite{fru_huber_2011}), we feel convinced to re-adjust
the battle, tentatively to 1310/11 BCE.
The subsequent island-hopping would lead to exactly our result for
Akhenaten.

\item
The radiocarbon-based analysis supplies another scientific method,
but it is has to cope with considerable errors like calibration
and refinement.
The year ``1280$\pm$12'' in our Table \ref{tab:kadesh} is simply
derived from the model by Ramsey \textit{etal}
\cite{fru_ramsey_2010} for the accession of Ramesses II on the
95\% confidence level:
take the medium of 1297 and 1273, as given in that paper, then
subtract 5 years for the battle.
The authors admit that their average calendrical precision is off
by 24 years for the New Kingdom and also recommend earlier years
than the current consensus.

\end{enumerate}

Astronomy returns the most substantial arguments for dates, but
it serves as an assistant only.
It cannot replace the adjustment of the historical plot into the
global picture.
The development of a story long ago has to be re-constructed by
historians.
Within the setting we suggest the year 1382 BCE for Akhenaten's
accession.
Our concept of three eclipses (1399, 1389, and 1378 BCE) is based
on plausibility, but we do not insist on having found the ultimate
solution.

The timeline was already put forward by others in a similar way
deploying astronomical dating, and we hope to support this view by
the method of eclipses.
That ``long'' chronology is reluctantly accepted by historians who
stick to their conventional king counts.
If they are willing to accept the reason for Akhenaten's venture,
as brought forward in this paper, re-arrangements would need to be
made in the lengths of reigns and probably some re-interpretation
of the textual evidence.


\section*{Acknowledgements}

This work is part of the Habilitation submitted to the University
of Heidelberg, Germany, in February 2020 \cite{finsternisbuch}.
The present version of the paper is overhauled, linguistically
improved, and now published ``as is''.
The entire research was performed in dire straits.

\newpage


\end{document}